\documentclass[reprint,prl,author-numerical]{revtex4-2}
\pdfoutput=1

\usepackage{graphicx}
\usepackage{dcolumn}
\usepackage{hyperref} 
\usepackage{mathtools}
\usepackage{siunitx}

\usepackage{graphicx}
\usepackage{float}
\usepackage{bm}
\usepackage{color}
\usepackage[dvipsnames]{xcolor}

\renewcommand{\vec}[1]{\mathbf{#1}}

\begin{document}
	
\title{Dielectric properties of aqueous electrolytes at the nanoscale}

\author{Maximilian R. Becker}
\affiliation{Fachbereich Physik, Freie Universität Berlin, Arnimallee 14, Berlin, 14195, Germany}

\author{Philip Loche}
\affiliation{Laboratory of Computational Science and Modeling, IMX, École Polytechnique Fédérale de Lausanne, 1015 Lausanne, Switzerland}
\affiliation{Fachbereich Physik, Freie Universität Berlin, Arnimallee 14, Berlin, 14195, Germany}

\author{ Roland R. Netz}
\email[]{rnetz@physik.fu-berlin.de}
\affiliation{Fachbereich Physik, Freie Universität Berlin, Arnimallee 14, Berlin, 14195, Germany}

\author{Douwe Jan Bonthuis}
\affiliation{Institute of Theoretical and Computational Physics, Graz University of Technology, Graz, Austria}

\author{Dominique Mouhanna}
\affiliation{Sorbonne Universit{\'e}, CNRS, Laboratoire de Physique Th{\'e}orique de la Mati{\`e}re Condens{\'e}e (LPTMC, UMR 7600), F-75005 Paris, France}

\author{H{\'e}l{\`e}ne Berthoumieux}
\email[]{helene.berthoumieux@espci.fr}
\affiliation{Gulliver, CNRS, {\'E}cole Sup{\'e}rieure de Physique et Chimie Industrielles de Paris, Paris Sciences et Lettres Research University, Paris 75005, France}

\begin{abstract}
Despite the ubiquity of aqueous electrolytes, the effect of salt on water organization remains controversial. We introduce a nonlocal and nonlinear field theory for the nanoscale polarization of ions and water and derive the electrolyte dielectric response as a function of salt concentration to first order in a loop expansion. By comparison with molecular dynamics simulations, we show that rising salt concentration induces a dielectric permittivity decrement and Debye screening in the longitudinal susceptibility but leaves the water structure remarkably unchanged. 
\end{abstract}

\maketitle

{\it Introduction - }
Studying electrolytes at the nanoscale is exciting for both the experimental relevance and the theoretical challenge~\cite{bjorneholm16,kavokine20,backus21}.
The nanometer scale is typical of technological and biological devices, it is the screening length of medium-concentrated
ionic solutions, as well as the range at which water
starts to behave as a discrete molecular medium~\cite{schlaich2016,fumagalli2018,monetprl}.  In the standard Poisson-Boltzmann (PB) approach, water is described as a linear dielectric medium, and its permittivity $\epsilon_w$ is wavenumber independent. This model cannot capture the complexity of water-ion interaction at the nanoscale.
A nonlocal dielectric medium is characterized by the wavenumber dependence of the susceptibility tensor $\chi(\vec{q})$, where $q$ is the scattering wavenumber, $q=2\pi/\lambda$. This kernel is defined by the correlations of the polarization field $\bm{\mathcal{P}}$ as ${\chi_{ij}(\vec{q})=\beta\langle \bm{\mathcal{P}}_i(\vec{q})\cdot\bm{\mathcal{P}}_j(-\vec{q})\rangle/\epsilon_0}$, with $\beta$ the inverse temperature and $\epsilon_0$ the vacuum permittivity~\cite{bopp1996,kornyshevtrans}.\par
\begin{figure}
	\includegraphics[scale=1]{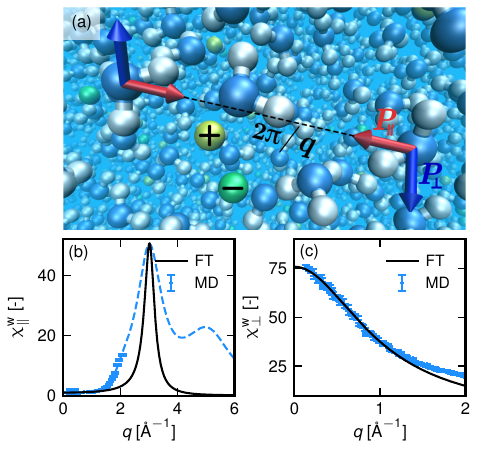} 
	\caption{a) Sketch of the system under study. Water acts as a nonlocal, nonlinear dielectric medium in the presence of ions. Using FT, we  evaluate the ionic-strength effect on the water
		longitudinal $\bm{\mathcal{P}_\parallel}$ (red arrow) and transverse $\bm{\mathcal{P}_\perp}$ (blue arrow) polarization correlations and compare with simulated response functions. Longitudinal (b) and transverse (c) MD simulated and FT-derived susceptibilities for pure water in Fourier space.   FT predictions follow from Eq.~(\ref{chiw}) for $K$= 1/76,  $\kappa_l$= -0.218~\AA$^{2}$ , $\alpha$=0.012~\AA$^{4}$ and $\kappa_t$=0.013~\AA$^{2}$. MD simulated susceptibilities are calculated from the charge structure factor  from radial distribution functions for larger $q$ (dashed line) and for discrete wavenumbers at low $q$ (data points) with a minimal $q=2\pi/L\approx$0.1\AA$^{-1}$ fixed by the size $L$ of the simulation box~\cite{sm_paper}. 
	}
	\label{fig:0}
\end{figure}

 For pure water, figure \ref{fig:0} shows the longitudinal $\chi^{\rm w}_{\parallel}$ and transverse $\chi^{\rm w}_\perp$  susceptibilities as a function of $q$ (panel b and c, blue markers and blue broken line) derived from force-field molecular dynamics (MD) simulations with the TIP4p/$\epsilon$ water model~\cite{sm_paper,gromacs,neuman83,flyvbjerg1989}, designed to reproduce its macroscopic permittivity~\cite{azcatl2014}. 
MD data display a pronounced maximum with $\chi^{\rm w}_{\parallel}(q)\gg$1 around $q$=3~\AA$^{-1}$, which corresponds to a $q$-region associated with a negative permittivity, as $\epsilon(q)$ obeys $\epsilon(q)=1/(1-\chi^{\rm w}_{\parallel}(q))$, named overscreening~\cite{bopp1996}. This phenomenon can be attributed to the short-range H-bond network of water~\cite{kornyshev1997}. 
	The transverse susceptibility is rather featureless and decays monotonically. \par
	Adding ions induces a decrement of the macroscopic permittivity $\epsilon(q=0)$ of electrolytes which is well-documented experimentally~\cite{hasted48}.  This effect can be recovered theoretically using a nonlinear PB equation~\cite{levy2012, levy2013}, which accounts for the low permittivity due to the water orientation in the hydration shells. \par
	The description of water-ion interactions has been the subject of many experimental~\cite{omta,chen2016,zhang2022,balos2022} and simulation works~\cite{galli2017,pluharova2017}. However, the effect of salt on the H-bond network - breaking or reinforcing  -  and the spatial range on which it plays a role remains controversial.  Coarse-grained theories that account for nonlocal dielectric properties, {\it ie} the $q$-dependence of the susceptibility,  provide a useful framework for describing correlated  fluids~\cite{kornyshev1986,Basilevsky1998,hildebrandt2004,maggs2006}. They have been applied to electrolytes to study the coupling between the correlation length of simple fluids and the Debye screening length by deriving a nonlocal linear PB equation~\cite{paillusson2010,benyaakov2011}. Recently, a general theory for electrolytes, including electrostatic and structural interactions for the solvent, has been derived ~\cite{blossey,blossey2,blossey3}. However, a nonlocal field theory (FT) for aqueous electrolytes, including the water-salt interaction at the nanoscale, validated by MD simulations, is missing. This is what we address in this letter. \par
{\it Nonlocal model for pure water - } 
 The electrostatic energy of pure water $\mathcal{U}_{\rm el}$ can be written as a functional of the polarization~\cite{blossey} $\bm{\mathcal{P}}(\vec{r})$ as:
\begin{eqnarray}
\label{Hframework}
\mathcal{U}_{\rm el}[\bm{\mathcal{P}}]=\frac{1}{2}\int d\vec{r} d\vec{r}' \frac{ \nabla\cdot \bm{ \mathcal{P}}(\vec{r}) \nabla\cdot \bm{ \mathcal{P}}(\vec{r}')}{4\pi \epsilon_0|\vec{r}-\vec{r}'|} 
+\mathcal{U}_{\rm conf}[\bm{\mathcal{P}}].
\end{eqnarray}
The first term corresponds to the bare Coulomb interactions between the partial charges $-\nabla\cdot\bm{\mathcal{P}}(\vec{r})$ of the fluid and the second term to a phenomenological configurational energy of the fluid~\cite{maggs2006,blossey} written as follows:
\begin{eqnarray}
\label{Hpw}
\mathcal{U}_{\rm conf}[\bm{\mathcal{P}}]&=&\frac{1}{2\epsilon_0}\int d\vec{r}\Big [\gamma \bm{ \mathcal{P}}(\bm{r})^4+ K \bm{ \mathcal{P}}(\bm{r})^2
+\kappa_l(\bm{\nabla}\cdot \bm{ \mathcal{P}}(\bm{r}))^2\nonumber\\&+&\kappa_t(\bm{\nabla}\times \bm{ \mathcal{P}}(\bm{r}))^2+ \alpha (\bm{\nabla}
(\bm{\nabla} \cdot \bm{ \mathcal{P}}(\bm{r}))  )^2 \Big ].
\end{eqnarray}
First, we discuss this nonlocal model at the Gaussian level, obtained by setting $\gamma$=0. In this case,  $\mathcal{U}_{\rm conf}[\bm{\mathcal{P}}]$ is a Landau-Ginzburg expansion up to the second spatial derivative for the longitudinal part (terms in $\kappa_l$ and $\alpha$) and up to the first spatial derivative for the transverse part (term in $\kappa_t$). This functional captures the main features of dielectric properties of water at the nanoscale~\cite{maggs2006,berthoumieux2015,monetprl}.  
The polarization susceptibility $\chi^{\rm w}$ is obtained by inversion of Eq.~(\ref{Hframework})~\cite{sm_paper}. We decompose it in Fourier space into a longitudinal $\chi_{\parallel}^{\rm w}$ and a transverse $\chi_{\perp}^{\rm w}$ response using translational invariance of the system, such that $\chi_{ij}(\vec{q})=\chi^{\rm w}_{\parallel}(q) q_iq_j/q^2+\chi^{\rm w}_{\perp}(q)(\delta_{ij}-q_iq_j/q^2)$, with $(i,j)=(x,y,z)$.  Their expressions follow from Eq.~(\ref{Hpw}) as 
\begin{equation}
\label{chiw}
	\chi_{\parallel}^{\rm w}(q)=\frac{1}{1+K+\kappa_l q^2+ \alpha q^4}, \quad \chi_{\perp}^{\rm w}(q)=\frac{1}{K+\kappa_t q^2}.
\end{equation}
For $\chi_\parallel^{\rm w}(q)$, the case $(\kappa_l<0$, $\alpha>0)$, generates a maximum at finite $q$. First, we determine the the parameters  ($K$, $\kappa_l$, $\alpha$) by fitting $\chi^{\rm w}_{\parallel}(0)$, the position and the value of the maximum of $\chi^{\rm w}_{\parallel}(q)$ to the MD data shown in Fig.\ref{fig:0} (b). Note that the secondary peak observed for $q=5$~\AA$^{-1}$ corresponds to intramolecular correlations and to length scales that are not addressed in our formula of $\mathcal{U}_{\rm conf}$.
Second, we determine  $\kappa_t$ by fitting the decay of $\chi_{\perp}^{\rm w}(q)$ at low $q$ shown in the MD data in Fig.~\ref{fig:0} (c). The values of $(K,\kappa_l,\alpha,\kappa_t)$ are given in the caption of Fig.~\ref{fig:0}.

For $\gamma> 0$, the term scaling like $\bm{\mathcal{P}}^4$ in Eq.~(\ref{Hpw}) describes the nonlinear water response. As a result, when submitted to a strong electrostatic field $\vec{E}$, the polarization no longer scales linearly with  $\vec{E}$ but as $\vec{E}^{1/3}$   ~\cite{alper90,fedorov2007,levy2013,berthoumieux2019}.
This accounts for the polarization of water in the ionic solvation shells~\cite{vorotyntsev2019,berthoumieux2019} and produces a threshold polarization, $P_0=\sqrt{K/2\gamma}$, between a linear-response  ($|\bm{\mathcal{P}}|\ll P_0$) and a saturation-response ($|\bm{\mathcal{P}}|\gg P_0$) regime. The parameter $\gamma$ tunes this saturation threshold.\par
In this letter, we compute the free energy and the nonlocal dielectric susceptibility of an aqueous electrolyte for which water is modeled by Eq.~(\ref{Hpw}). We perform MD simulations and compare the longitudinal and transverse permittivity dependence for varying salt concentrations with the FT predictions. Finally, we identify the essential building blocks of a FT that accurately models electrolytes at the nanoscale and reproduces the longitudinal
and transverse dielectric susceptibilities computed by MD simulations. \par
{\it Nonlocal model for electrolytes - } We consider an electrolyte with $N$ point-like cations of charge $e$ and $N$ point-like anions of charge $-e$ solvated in water. The ionic charge density reads $\rho(\vec{r})=\Sigma_{i=1}^{N} e \delta(\vec{r}-\vec{r}_i^+)-\Sigma_{j=1}^{N} e \delta(\vec{r}-\vec{r}_j^-)$.
In the canonical ensemble, the partition function of the electrolyte can be written as 
\begin{eqnarray}
\label{Zloc}
&&\mathcal{Z}=\frac{1}{(N!)^2}\left[\prod_{i=1}^{N}\int d\vec{r}_i^+ \right]\left[\prod_{j=1}^{N}\int d\vec{r}_j^- \right]\nonumber\\& &\times\int \mathcal{D}[\bm{\mathcal{P}}]e^{-\beta \mathcal{U}_{\rm conf}[\bm{ \mathcal{P}}]}
 e^{-\frac{\beta}{2}\int d\vec{r} d\vec{r}'\rho_{\rm tot}(\vec{r})v(\vec{r}-\vec{r}')\rho_{\rm tot}(\vec{r}')}\,
\end{eqnarray}
with the total charge density $\rho_{\rm tot}(\vec{r})=\rho(\vec{r})-\nabla\cdot\bm{\mathcal{P}}(\vec{r}) $ and $v(\vec{r})=1/4\pi \epsilon_0|\vec{r}|$. $\mathcal{Z}$ includes the configurational degrees of freedom of the polarizable water, with $\int\mathcal{D}[\bm{\mathcal{P}}]$ the path integral over the fluctuating field $\bm{\mathcal{P}}$, and the Coulomb interactions between free and partial charges. Following the work of Orland and co-workers~\cite{levy2012}, we perform a Hubbard-Stratonovich transformation for $\mathcal{Z}$ to get rid of the long-range potential $v$ and switch to the grand-canonical ensemble for a more tractable expression of the partition function. The grand-canonical partition function, ${\Xi=\int \mathcal{D}[\bm{\mathcal{P}}]\, \mathcal{D}[\Psi]e^{ - \beta F_u[\bm{{\mathcal{P}}},\Psi]}}$, takes in the Gaussian limit ( $\gamma=0$ ) a simple form as a function of the action $F_u[{\bm {\mathcal P}},\Psi]$ which is a functional of ${\bm {\mathcal P}}$ and of the electrostatic potential $\Psi$~\cite{levy2012}. Assuming a 1:1 electrolyte and an ionic density $n=N/V$ with $V$ the volume of the system, we find:
\begin{eqnarray}
\label{action}
F_u[{\bm {\mathcal P}},\Psi]&=& \mathcal{U}_{\rm conf}[\bm{\mathcal{P}}] -\int d\vec{r}\Big(\frac{\epsilon_0}{2} (\nabla \Psi(\vec{r}))^2-\Psi(\vec{r}) \nabla \cdot \bm{\mathcal{P}}(\vec{r})\nonumber\\&-&\frac{2 n}{\beta}\cosh(\beta \Psi(\vec{r}) e)\Big).
\end{eqnarray}
Details of the calculations are given in the subsection S2.2 of the SM. The free energy $\mathcal{F}$ of the system follows from $\mathcal{F}=-k_BT\rm{ln} \Xi$ and the generalized susceptibility is a 4$\times$4 matrix defined as:
\begin{eqnarray}
\label{DefSusc}
&&\left(\begin{array}{cc} \epsilon_0\chi & \chi_{P,\psi} \\ \chi_{\psi, P} & \chi_{\psi,\psi}/\epsilon_0
	\end{array}\right)({\vec{r}_1-\vec{r}_2})= \nonumber\\ & &\left(\begin{array}{cc}
\frac{\delta^2 F_u(\bm{P},\psi)}{\delta \mathcal{P}_i(\vec{r}_1) \delta \mathcal{P}_j(\vec{r}_2) }   & \frac{\delta^2 F_u(\bm{P},\psi)}{\delta \mathcal{P}_i(\vec{r}_1) \delta \Psi(\vec{r}_1)} \\
\frac{\delta^2 F_u(\bm{P},\psi)}{\delta \Psi(\vec{r}_1) \delta \mathcal{P}_j(\vec{r}_2)}  & \frac{\delta^2 F_u(\bm{P},\psi)}{\delta \Psi(\vec{r}_1) \delta \Psi(\vec{r}_2)}
\end{array}\right) ^{-1}
\end{eqnarray}
where the second functional derivatives of $F_u[{\bm {\mathcal P}},\Psi]$ are evaluated at the mean field point 
($\bm{P},\psi$), defined by $\delta F_u(\bm{P},\psi)/\delta{\mathcal{P}_i}=\delta F_u(\bm{P},\psi)/\delta \psi=0$~\cite{sm_paper}.  We focus here on the polarization susceptibility. The longitudinal part is written in Fourier space as
\begin{figure}
\includegraphics[scale=0.9]{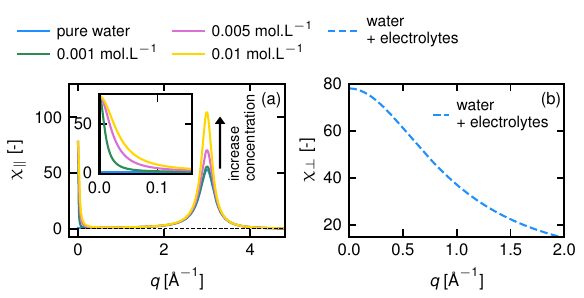} 
\caption{Dielectric susceptibility for electrolytes. (a) Longitudinal susceptibility $\chi_{\parallel}$ - Eq.~(\ref*{SuscepElectMT}) - as a function of $q$ for water and solutions of increasing salt concentration. The arrow indicates the increase of the peak maximum with an increase of the concentration. The inset zooms into the low $q$ part of the plot. (b) Transverse susceptibility $\chi_{\perp}=\chi_{\perp}^{\rm w}$ given by Eq~(\ref{chiw}) as a function of $q$, which is identical for water and electrolytic solutions. The parameter values of the FT
	model given in Eq. (\ref{Hpw}) are $K$= 1/76,  $\kappa_l$= -0.218~\AA$^{2}$, $\alpha$=0.012~\AA$^{4}$ and $\kappa_t$=0.013~\AA$^{2}$.}
\label{fig:1}
\end{figure}

\begin{equation}
\label{SuscepElectMT}
\chi_{\parallel}(q)=\frac{\frac{\epsilon_w}{\lambda_D^2}+q^2}{\left(\frac{\epsilon_w}{\lambda_D^2}+q^2\right)(K+\kappa_l q^2+\alpha q^4)+q^2}, 
\end{equation}
with the Debye length $\lambda_D=\sqrt{\epsilon_0\epsilon_w/2\beta n e^2}$~\cite{sm_paper}. Fig.~\ref{fig:1} a) shows $\chi_{\parallel}(q)$ for increasing molar salt concentration $c=n/\mathcal{N}_a$, where $\mathcal{N}_a$ is the Avogadro number.  The inset zooms on the low $q$ part of $\chi_{\parallel}(q)$. It reveals that for low $q$, $\chi_{\parallel}(q)$ decreases with a characteristic wavenumber $q^{\rm char}$, which scales as $1/\lambda_D$. The Gaussian model predicts an enhancement of the water ordering with an increase of $c$, as indicated by the magnitude increase of the peak at $q$=3~\AA$^{-1}$. This can be understood as follows: in the nonlocal Gaussian framework, an ion, located at $r=0$, generates an oscillating exponentially decaying polarization response~\cite{Basilevsky1998,vatin21}. This induces the organization of the free ionic charges, which in turn generates longer range correlations in water that increase with the salt concentration until a nonphysical crystallization of the system occurs, corresponding to a divergence of $\chi_{\parallel}(q)$ (see Fig. S2 of the SM). In contrast, for a very dilute solution, the "overscreening" peak at $q$=3~\AA$^{-1}$ is identical to the peak of neat water as shown by the green line in fig.~(\ref{fig:1}) a).  In this case, the Debye length is several orders of magnitude larger than the polarization correlation lengths in water. $\chi_{\parallel}(q)$  in Eq.~(\ref{SuscepElectMT}) takes a simple expression in the low $q$ regime,  $\chi_{\parallel}(q)\approx(\frac{\epsilon_w}{\lambda_D^2}+q^2)/(K\left(q^2+\frac{\epsilon_w}{\lambda_D^2}\right)+q^2)$, and becomes a function of only $\lambda_D$ and the large-scale ($q=0$) parameter $K$. In the large $q$ regime,   $\chi_{\parallel}(q)$ equals the neat water  susceptibility, $\chi_{\parallel}(q)\approx \chi_\parallel^w(q)$ in Eq.~(\ref{chiw}). Up to a negligible constant, we can write $\chi_{\parallel}(q)$ for dilute electrolytes as:
 \begin{equation}
 \label{decoupled_chi}
 	 \chi_{\parallel}(q)\approx\frac{\frac{\epsilon_w}{\lambda_D^2}+q^2}{K\left(q^2+\frac{\epsilon_w}{\lambda_D^2}\right)+q^2}+\frac{1}{1+K+\kappa_l q^2+ \alpha q^4}.
 \end{equation}
The transverse susceptibility $\chi_\perp(q)$ of electrolytes is unaffected by the presence of salt and obeys $\chi_{\perp}(q)$=$\chi_\perp^{\rm w}(q)$, see Fig.~\ref{fig:1} (b). Indeed, the coupling between the salt and the solvent occurs via the Coulomb interactions and involves only the longitudinal part of the polarization, as seen in Eq.~(\ref{Zloc}). Finally, we note that the Gaussian model predicts that the dielectric large-scale properties of electrolyte solutions $\chi_{\parallel}(0)=\chi_{\perp}(0)=1/K$ are independent of the salt concentration, which contradicts the permittivity decrement due to added salt~\cite{hasted48}, and is a consequence of the simple Gaussian model with $\gamma$=0.\par
\begin{figure}
\includegraphics[scale=0.9]{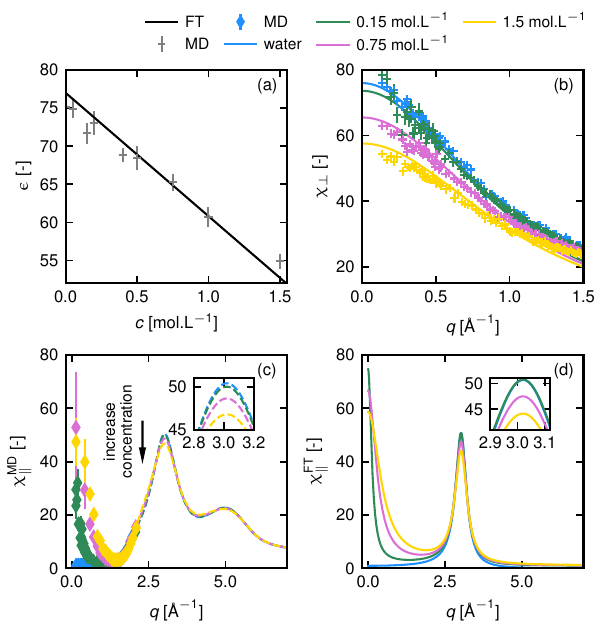} 
\caption{Response functions of electrolytes. (a)  MD simulated (markers) and FT-derived (line) permittivity as a function of salt concentration $c$. FT expression is given in Eq.~(\ref{epsoneloop}) for $\delta K_c$=0.0028~mol$^{-1}$.L. (b) MD simulated (markers) and FT-derived (line, Eq~(\ref{chiperpFT}) transverse permittivity. (c) MD simulated longitudinal susceptibility $\chi_{\parallel}^{\rm MD}$.  Markers and dashed lines correspond to two extraction methods (see caption of Fig.{\ref{fig:0}} and S1 of the SM). (d) FT expression of the longitudinal susceptibility $\chi_{\parallel}^{\rm FT}$ given in Eq.~(\ref{chiparallelFT}). The insets zoom on the $q$=3\AA$^{-1}$ peak. For (d), the blue and green plots overlap completely.} 
\label{fig:2}
\end{figure}

%
{\it Comparison with MD simulations - }
To check the validity of the Gaussian model, we compare its predictions with the dielectric properties of simulated solutions of NaCl in TIP4p/$\epsilon$ water for concentrations $c$ up to 1.5~mol.L$^{-1}$~\cite{azcatl2014,loche2021}.  We compute the $q=0$ permittivity and plot $\epsilon(c)$ in Fig.~\ref{fig:2}~(a). We observe a linear decay not described by the Gaussian model, which predicts a constant $q$=0 permittivity. Panels (b) and (c) of Fig.~\ref{fig:2} show $\chi_{\perp}^{\rm MD}(q)$ and  $\chi_{\parallel}^{\rm MD}(q)$ for $c$=0.15~mol.L$^{-1}$ ($\lambda_D$=7.8~\AA), $c$=0.75~mol.L$^{-1}$ ($\lambda_D$=3.5~\AA), and $c$=1.5~mol.L$^{-1}$ ($\lambda_D$=2.5~\AA). The blue markers and blue broken line show the response for pure water.  $\chi^{\rm MD}_{\perp}(q\rightarrow 0)$ decays for an increasing salt concentration. $\chi^{\rm MD}_{\parallel}(q)$ shows a decay at low $q$  ($q\leq$2~\AA $^{-1}$) associated with a cut-off wavenumber of about $1/\lambda_D$. At large $q$ ($q>$2~\AA$^{-1}$), $\chi_{\parallel}^{\rm MD}(q)$ for electrolytes surprisingly remains almost identical to the one of pure water. It is only at high salt concentration (yellow markers and line, $c$=1.5~mol.L$^{-1}$) that an effect of the salt is visible: the peak at $q\approx$3~\AA$^{-1}$ slightly decreases and flattens - see inset of Fig.~\ref{fig:2} (c), in contradiction to the Gaussian model predictions. \par
{\it Nonlinear model for water - } 
To obtain a better agreement between FT and MD simulations, we consider the nonlinear configuration energy  $\mathcal{U}_{\rm conf}$ with $\gamma > 0$. The action $F_u$ contains then a non-quadratic term, $\gamma\bm{{\mathcal{P}}}^4$. Its effect on response functions can be estimated by using a {\it loop expansion} around the mean field~\cite{netz2000} that we perform here at first order~\cite{levy2013}.  $F_u$ is expanded up to the second order in $(\bm{ \mathcal{P}},\Psi)$ around the mean field solution according to $F_u[\bm{\mathcal{P}},\Psi]\approx F_u[{\bm P},\psi] +1/2\int d\vec{r}d\vec{r}'(\delta \bm{P}(\vec{r}),\delta \psi(\vec{r}))\cdot F_u^{(2)}({\bm P}, \psi)\cdot (\delta \bm{P}(\vec{r'}),\delta \psi(\vec{r}'))$, with $F_u^{(2)}$ the second functional derivative of $F_u$ with respect to $(\bm{\mathcal{P}}, \Psi)$ and $\delta \bm{P}=\bm{\mathcal{P}}-{\bm P}$, $\delta \psi=\Psi-\psi $. The partition function thus follows as 
\begin{eqnarray}	
	 \Xi\approx \exp \left\{- \beta F_u[\bm{{\bm{P}}},\psi]-\frac{1}{2} \ln \left[\det \beta F_u^{(2)}({\bm P}, \psi)\right]\right\}
	 \end{eqnarray}
and the free energy is written  as 
$\mathcal{F}\approx F_u[{\bm P},\psi]+{1/2}{\rm Tr}{\rm ln}(\beta F_u^{(2)}[{\bm P},\psi])/\beta$~\cite{sm_paper}. The inverse susceptibility follows as $
\chi^{-1}(\vec{r}_1-\vec{r}_2)=\chi^{-1}_{(0)}(\vec{r}_1-\vec{r}_2)+\chi^{-1}_{(1)}(\vec{r}_1-\vec{r}_2)$, with  $\chi_{(0)}^{-1}$ the inverse susceptibility in the Gaussian limit given in Eq.~(\ref{DefSusc}). The one-loop correction term $\chi^{-1}_{(1)}(\vec{r}_1-\vec{r}_2)$ obeys:
\begin{eqnarray}
\label{chifirstloop}
(\chi^{-1}_{(1)})_{i,j}(\vec{r}_1-\vec{r}_2)=\frac{1}{2\beta}\frac{\delta ^2 {\rm Tr}\ln \beta F_u^{(2)}({\bm P},\psi)}{\delta \mathcal{P}_i(\vec{r}_1) \delta \mathcal{P}_j(\vec{r}_2)}.
\end{eqnarray}
 Performing the functional derivatives and calculating the trace in Eq.~(\ref{chifirstloop}), one obtains~\cite{sm_paper}
\begin{eqnarray}
\label{chim11}
&&(\chi_{(1)}^{-1})_{i,j}(\vec{r}_1-\vec{r}_2)=\delta K \delta(\vec{r}_1-\vec{r}_2)\delta_{ij}\quad {\rm with}\nonumber\\
\delta K&=&\frac{20\gamma \epsilon_0}{3 \beta }\left( \hat{\chi}_{\parallel}(r=0)+2\hat{\chi}_{\perp}(r=0)\right).
\end{eqnarray}
Note that here $\hat{\chi}_\parallel$ and $\hat{\chi}_\perp$  are  back-Fourier transforms of Eqs.~(\ref{chiw}, \ref{SuscepElectMT}) at $r=0$.
 The first-order correction of the susceptibility is purely local and depends on $c$ via $\chi_\parallel$.
We expand $\epsilon=1+1/(K+\delta K)$ linearly in the salt concentration $c$ and obtain: 
\begin{eqnarray}
\label{epsoneloop}
	\epsilon(c)&=&\epsilon_w-\frac{\delta K_c}{K^2}c.
\end{eqnarray}
with $\delta K_c$ given in S2.8 of the SM.
We determine $\delta K_c$ by fitting the permittivity of simulated solutions as shown in fig.~\ref{fig:2} (a). The value of $\delta K_c$ is given in the caption of Fig.\ref{fig:2}.\par

{\it Susceptibility kernels for electrolytes - } We now compare the simulated $q$-dependent susceptibilities with the nonlinear FT ones, obtained by replacing $K$ by ${K + \delta K_c c}$ in Eq.~(\ref*{SuscepElectMT}) for $\chi_{\parallel}$ and in Eq.~(\ref{chiw}) for $\chi_{\perp}$, and using the values $(\alpha, \kappa_l, \kappa_t, K)$ fitted for pure water. 
We plot the one-loop corrected transverse response,
\begin{equation}
\label{chiperpFT}
\chi^{\rm FT }_\perp(q)=\frac{1}{K+\delta K_c c +\kappa_t q^2},
\end{equation}
in Fig.~\ref{fig:2} (b)  for $c$=0, 0.15, 0.75, 1.5 mol.L$^{-1}$ and find a very reasonable
agreement with simulations. \par In the SM (see Fig. S4), we show that the renormalized longitudinal susceptibility exhibits an enhancement of the "overscreening" peak, which is weaker than for the Gaussian limit but still contradictory to the MD data. To account for the structural decoupling of water and ion polarization, we use
the decoupled expression for $\chi_\parallel(q)$ given in Eq.~(\ref{decoupled_chi}) and replace $K$ by $K+\delta K_cc$. We obtain:
\begin{eqnarray}
\label{chiparallelFT}
\chi^{\rm FT }_\parallel(q)&=&\frac{\frac{\epsilon_w}{\lambda_D^2}+q^2}{(K+\delta K_c c)(\frac{\epsilon_w}{\lambda_D^2}+q^2)+q^2}\nonumber\\&+&\frac{1}{1+K+\delta K_cc+\kappa_l q^2+\alpha q^4}.
\end{eqnarray}
The first term accounts for Debye screening, the second term for the susceptibility for pure water. In both terms, the $q=0$ parameter $K$ is renormalized.  We plot $\chi^{\rm FT }_\parallel(q)$ for $c$=0, 0.15, 0.75, 1.5 mol.L$^{-1}$ in Fig.~\ref{fig:2} (d). It reproduces the MD main features, in particular, the decay at low $q$ and the flattening of the peak at $q$=3~\AA$^{-1}$ with an increase of the salt concentration.\par
{\it Discussion - }
In this letter, we describe the behavior of the susceptibility of water with added salt.  For this, we compare the susceptibility calculated from FT, including nonlocal and nonlinear terms, with MD data. We highlight two regimes in $q$-space for the longitudinal correlations: for small $q$, $q\leq$2~\AA$^{-1}$, the electrolyte is described with a renormalized permittivity $\epsilon(c)$. This long-range interaction regime shows the standard Debye screening. At larger $q$, the longitudinal susceptibility in electrolytes equals the one of pure water but is associated with the renormalized permittivity $\epsilon(c)$. 
These two decoupled regimes reflect two distinct water structures in electrolytes: water molecules solvating ions are "electrically" frozen by the ionic field, creating a solvation shell of vanishing permittivity~\cite{levy2012}. Outside of this shell, the water structure is neither strengthened nor destroyed, but is remarkably unaffected.~\cite{monetJCP2021,zhang2022}. 
Our work reveals the absence of coupling between salt screening and transverse	polarization modes of water, which could have consequences on the interactions between objects immersed in electrolytes~\cite{pires21} that are assumed to be screened~\cite{MahantyNinham1976,Parsegian2006}.
This study gives a clear picture of the nature and range of salt's effect on water organization at the nanoscale for unconfined solutions. It is the first step toward an accurate FT describing the properties of
nanoconfined electrolytes~\cite{tuladhar2020,robert23,martinjimenez2016,lee2021}.\\

HB thanks H. Orland for fruitful discussions. H.B. acknowledges funding from the Humboldt Research Fellowship Program for Experienced Researchers.  M.B. acknowledges support by Deutsche Forschungsgemeinschaft, Grant No. CRC 1349, Code No. 387284271, Project No. C04. This work was funded by the Deutsche Forschungsgemeinschaft (DFG, German Research Foundation) in project GRK 2662 -  434130070.

\bibliographystyle{unsrt}
\bibliography{RPfinal}
\end{document}


\title{Supplementary Materials for\\Dielectric properties of aqueous electrolytes at the nanoscale}
	
\author{Maximilian R. Becker}
\affiliation{Fachbereich Physik, Freie Universität Berlin, Arnimallee 14, Berlin, 14195, Germany}

\author{Philip Loche}
\affiliation{Laboratory of Computational Science and Modeling, IMX, École Polytechnique Fédérale de Lausanne, 1015 Lausanne, Switzerland}
\affiliation{Fachbereich Physik, Freie Universität Berlin, Arnimallee 14, Berlin, 14195, Germany}

\author{ Roland R. Netz}
\email[]{rnetz@physik.fu-berlin.de}
\affiliation{Fachbereich Physik, Freie Universität Berlin, Arnimallee 14, Berlin, 14195, Germany}

\author{Douwe Jan Bonthuis}
\affiliation{Institute of Theoretical and Computational Physics, Graz University of Technology, Graz, Austria}

\author{Dominique Mouhanna}
\affiliation{Sorbonne Universit{\'e}, CNRS, Laboratoire de Physique Th{\'e}orique de la Mati{\`e}re Condens{\'e}e (LPTMC, UMR 7600), F-75005 Paris, France}

\author{H{\'e}l{\`e}ne Berthoumieux}
\email[]{helene.berthoumieux@espci.fr}
\affiliation{Gulliver, CNRS, {\'E}cole Sup{\'e}rieure de Physique et Chimie Industrielles de Paris, Paris Sciences et Lettres Research University, Paris 75005, France}

	\maketitle
	
	\tableofcontents

\section{Supplementary information for molecular dynamics simulation}

\begin{figure}
	\includegraphics[scale=0.6]{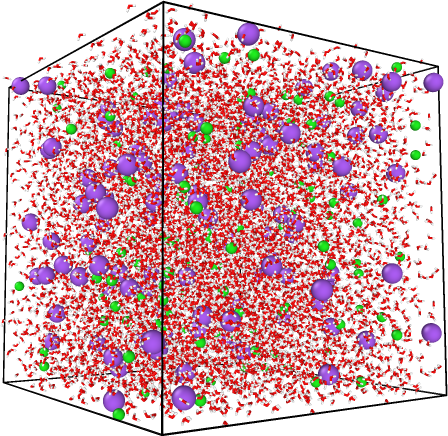}
	\caption{Snapshot of the simulation box of an electrolyte. Red and white sticks represent TIP4p/$\epsilon$ water molecules, purple spheres chloride Cl$^-$ ions, and green spheres sodium Na$^+$ ions. This picture corresponds to a 0.75 mol.l$^{-1}$ solution. Size of the box: $L$=6.5~nm.}
	\label{fig:5}
\end{figure}

\paragraph{Molecular dynamics simulations}
We simulate a cubic water box of side $L$=6.5 nm composed of N$_w$ water molecules, N$_w$ going from   9033 to 8527 for increasing salt concentration. See a snapshot of the simulated system in Fig.~\ref{fig:5}. The 0.15 mol.l$^{-1}$ solution contains 25 ion pairs, the 0.75 mol.l$^{-1}$ solution contains 124 ion pairs and the 1.5 mol.l$^{-1}$ solution, 248 ion pairs.
We perform simulations with the TIP4p/$\epsilon$~\cite{azcatl2014}, which contains 4-interaction sites,  three point-charges, and one Lennard-Jones reference site model. The Lennard-Jones (LJ) center is placed on the oxygen. Charges are placed on the hydrogen atoms, and an additional interaction site, M, carries the negative charge. This model is derived from the TIP4p model with parameters modified to reproduce the macroscopic permittivity of water. The ions (Na$^+$ and Cl$^-$) are treated according to the force field developed in the reference~\cite{loche2021}.\par
Simulations are performed using the GROMACS 2021 molecular dynamics simulation package~\cite{gromacs}, and the integration time steps are set to $\Delta t$=2 fs. Simulation boxes are periodically replicated in all directions, and long-range electrostatics are handled using the smooth particle mesh Ewald (SPME) technique. Lennard-Jones interactions are cut off at a distance $r_{\rm cut}$=0.9 nm.  A potential shift is used at the cut-off distance. All systems are coupled to a heat bath at 300 K using a v-rescale thermostat with a time constant of 0.5 ps. We use MDAnanlysis to treat the trajectories. After creating the simulation box, we perform a first energy minimization. We equilibrate the system in the NVT ensemble for 200 ps and afterward in the NPT ensemble for another 200 ps using a Berendsen barostat at 1 bar.
Production runs are performed in the NVT ensemble for 20 ns.

\paragraph{Statistical treatment}
For the longitudinal and the transverse susceptibility, error bars are derived following the reblocking method~\cite{flyvbjerg1989}.
For the $q=0$ permittivity, we cut the trajectory in 5 statistically independent blocks, compute the permittivity of each block,  estimate the sample variance $\sigma^2$ and define the error bar as $\sqrt{\sigma^2/5}$. \par

\paragraph{Macroscopic permittivity}
Permittivity is calculated from the total system dipole moment~\cite{neuman83} $\mathbf{M}$ 
according to:
\begin{equation}
{\epsilon} -1 = \frac{\left\langle \mathbf{M} \cdot \mathbf{M} \right\rangle - \left\langle \mathbf{M} \right\rangle \cdot \left\langle \mathbf{M} \right\rangle}{3\epsilon_0 k_B T V},
\end{equation}
as implemented in the GROMACS dipoles module. $\mathbf{M}$ is the volume integral of the polarization as $\mathbf{M}=\int_Vd\vec{r} \bm{\mathcal{P}(\vec{r})}$ and $V$ is the volume of the box. Note that the ion polarization is not taken into account.
\paragraph{q-dependent susceptibilities}
To compute the $q$-dependent susceptibilities, we use the fluctuation-dissipation theorem, relating the response functions to the polarization fluctuations as follows:
\begin{equation}
\label{FDT}
\chi_{ij}(\vec{q})=\frac{\langle \mathcal{P}_i(\vec{q})\cdot\mathcal{P}_j(-\vec{q})\rangle}{\epsilon_0 k_BT}.
\end{equation}

One can express the longitudinal susceptibility as a function of the charge structure factor $S(q)$ and obtains
\begin{equation}
\chi_\parallel(q)=\frac{S(q)}{q^2 \epsilon_0 k_BT}.
\end{equation}

The charge structure factor in Fourier space can be decomposed into an intramolecular and an intermolecular part,
\begin{equation}
S(q)=S_{\rm int}(q)+S_{\rm inter}(q)\,
\end{equation}
with $S_{\rm inter}(q)$ the intermolecular contribution
\begin{equation}
S_{\rm inter}(q)=4 n_w z^2 e^2\left(h_{\rm MM}(q)+h_{\rm HH}(q)-2h_{\rm HM}(q)\right)
\end{equation}
$z$ is the valency, $e$ being the elementary charge, $n_w$ the molecular density. $h_{\rm IJ}$ is the Fourier transform of $g_{\rm IJ}(r)-1$, $g_{\rm IJ}(r)$ being the radial distribution function associated with the atoms couple IJ. 
The intramolecular contribution can be written as
\begin{equation}
S_{\rm intra}(q)=4 n_w z^2 e^2\left(\frac{\sin(q d_{\rm HH})}{q d_{\rm HH}}-4\frac{\sin(q d_{\rm HM})}{q d_{\rm HM}}+3\right)
\end{equation}
where $ d_{\rm IJ}$ is the intramolecular distance between atoms I and J.

At low $q$, the numerical precision of $S(q)$ becomes pretty low as the function $h_{\rm IJ}(r)$ is obtained on a finite range imposed by the box size.  To solve this problem, we proceed as follows. 
For $q<$ 2.5~\AA$^{-1}$, we take into account the periodicity of the system, calculate the charge structure factor for discretized values of the wavenumber $q$, $q=2\pi/L\sqrt{n_x^2+n_y^2+n_z^2}$. We compute  directly the charge structure factor from the charge distribution $	\tilde{\rho}(q)$ in the Fourier space, 
\begin{eqnarray}
\label{rhoq}
\tilde{\rho}(q)&=&\sum_{i=1}^{Nw} ez  e^{i \vec{q} \cdot \vec{r}}\left(-2 e^{-i\vec{q}\cdot\vec{r}_{{\rm M},i}}+e^{-i\vec{q}\cdot\vec{r}_{{\rm H1},i}}+e^{-i\vec{q}\cdot\vec{r}_{{\rm H2},i}} \right)
\end{eqnarray}
where H$_{1,i}$ and H$_{2,i}$ stand for the two hydrogens and M$_i$ for the site carrying the negative charge of the molecule $i$. The charge structure factor $S(q)=\langle \tilde{\rho}(q) \tilde{\rho}(-q) \rangle/ V$ follows

\begin{eqnarray}
S(q) &=&\frac{2q_H^2}{V}\sum_{i,j, j\leq i}\Big(4 \cos(\vec{q}\cdot\vec{d}_{\rm MiMj})-2\cos(\vec{q}\cdot\vec{d}_{\rm MiH1j})-2\cos(\vec{q}\cdot\vec{d}_{\rm H1iMj})\nonumber\\&-&2\cos(\vec{q}\cdot\vec{d}_{\rm MiH2j})-2\cos(\vec{q}\cdot\vec{d}_{\rm H2iMj})+\cos(\vec{q}\cdot\vec{d}_{\rm H1iH1j})+\cos(\vec{q}\cdot\vec{d}_{\rm H2iH2j})\nonumber\\&+&\cos(\vec{q}\cdot\vec{d}_{\rm H1iH2j})+\cos(\vec{q}\cdot\vec{d}_{\rm H2iH1j})\Big)
\end{eqnarray}
where $q$ is a vector and $d_{AiAj}$ stands for $\vec{r}_{Ai}-\vec{r}_{Aj}$.

The transverse susceptibility is calculated following ref.~\cite{kornyshevtrans}.
The polarization of the medium in Fourier space
\begin{equation}
{\bf P}({\bf q})=\Sigma_j {\bf p}_j({\bf q})e^{-i {\bf q}\cdot  {\bf r}_j}
\end{equation}
can be written as a sum over the molecular polarization ${\bf p}_j({\bf q})$ of the molecule $j$, which reads as
\begin{equation}
{\bf p}_j({\bf q})=	\frac{1}{\sqrt{V}}\Sigma_\alpha \frac{e z_\alpha {\bf \delta r}_{\alpha j}}{i {\bf q}\cdot {\bf \delta r}_{\alpha j } }\left(1-e^{-i{\bf q}\cdot {\bf \delta r}_{\alpha j }}\right)
\end{equation}
with ${\bf \delta r}_{\alpha j } $ the distance between the charge $\alpha$ and the center of mass of the molecule and $V$ the volume of the simulation box. 
We then take the transverse part of the polarization ${\bf P}_\perp({\bf q})={\bf q}\times {\bf P}(q)/q$ and define the transverse susceptibility as
\begin{equation}
\label{chiperpMD}
\chi_\perp(q)=\frac{\langle{\bf P}_\perp({\bf q})\cdot{\bf P}_\perp({\bf -q})\rangle}{2k_BT\epsilon_0}.
\end{equation}

Note that we replace $\left(1-e^{-i{\bf q}\cdot {\bf \delta r}_{\alpha j }}\right)/i {\bf q}\cdot {\bf \delta r}_{\alpha j } $ by 1 for $ {\bf q}\cdot {\bf \delta r}_{\alpha j } <$ 10$^{-5}$ to prevent numerical errors.

\section{Supplementary information for the field theory model}
\subsection{Derivation of $\chi^{w}(\vec{q})$, the susceptibility of pure water}
\begin{figure}
		\includegraphics{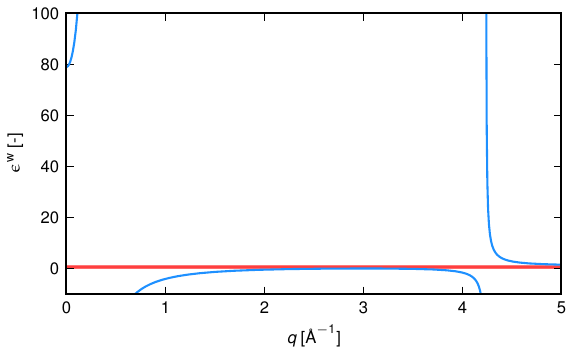}
		\caption{$q$-dependent permittivity of water. The permittivity $\epsilon^{\rm w}(q)$ is given in Eq.~(\ref{epsq}) and plotted as a function of $q$ for $K$= 1/76,  $\kappa_l$= -0.218~\AA$^{2}$, $\alpha$=0.012~\AA$^{4}$. The forbidden zone $[0,1[$ for $\epsilon^{\rm w}$~\cite{bopp1996} is shaded in red.}
		\label{fig:6}
\end{figure}

The dielectric susceptibility $\chi(\vec{r}-\vec{r}')$ of the system  described by Eq. (1) of the main text is defined as 

\begin{equation}\label{chiparaperp}
\begin{aligned}
	\mathcal{U}_{\rm el}[\bm{\mathcal{P}}]&=\frac{1}{2\epsilon_0}\int d\vec{r} d\vec{r}'\bm{\mathcal{P}}(\vec{r}) \cdot \chi^{w, -1}(\vec{r}-\vec{r}')\cdot \bm{\mathcal{P}}(\vec{r}'),\\
 &=\frac{1}{2\epsilon_0}\int d\vec{r} d\vec{r}'\left(\bm{\mathcal{P}}_\parallel(\vec{r}) \cdot\chi_\parallel^{w, -1}(\vec{r}-\vec{r}')\cdot \bm{\mathcal{P}}_\parallel(\vec{r}')+\bm{\mathcal{P}}_\perp(\vec{r}) \cdot\chi_\perp^{ w, -1}(\vec{r}-\vec{r}')\cdot \bm{\mathcal{P}}_\perp(\vec{r}')\right),
 \end{aligned}
\end{equation}
where we have split the polarization field into a longitudinal part $\bm{\mathcal{P}}_\parallel$ and a transverse part $\bm{\mathcal{P}}_\perp$, which respectively satisfy
$\bm{\nabla}_{\vec{r}}\times  \bm{\mathcal{P}}_\parallel(\vec{r})=0$ and $\bm{\nabla}_{\vec{r}} \cdot  \bm{\mathcal{P}}_\perp(\vec{r})=0$.
We rewrite the electrostatic energy as a functional of the longitudinal $\bm{\mathcal{P}}_\parallel(\vec{q})$ and transverse $\bm{\mathcal{P}}_\perp(\vec{q})$ polarization fields in Fourier space as
\begin{equation}
\begin{aligned}
	\mathcal{U}_{\rm el}[\bm{\mathcal{P}}]&=\frac{1}{2\epsilon_0}\int d\vec{q} \left(\bm{\mathcal{P}}_\parallel(\vec{q}) \cdot (1+K+\kappa_l q^2+\alpha q^4)\frac{q_iq_j}{q^2}\cdot \bm{\mathcal{P}}_\parallel(\vec{-\vec{q}})+\bm{\mathcal{P}}_\perp(\vec{q})\cdot (K+\kappa_t q^2)(\delta_{ij}-\frac{q_iq_j}{q^2} )\cdot \bm{\mathcal{P}}_\perp(-\vec{q})\right),
 \end{aligned}
\end{equation}
and invert the kernel $\chi^{-1,w}(\vec{q})$ to get Eq. (3) of the main text:
\begin{equation}
	\chi_{\parallel}^{\rm w}(q)=\frac{1}{1+K+\kappa_l q^2+ \alpha q^4}, \quad \chi_{\perp}^{\rm w}(q)=\frac{1}{K+\kappa_t q^2}.
\end{equation}
The $q$-dependent permittivity of the system obeys:
	\begin{equation}
	\label{epsq}
	\epsilon(q)=\frac{1}{1-\chi(q)}
	\end{equation}
	and is plotted in Fig.~(\ref{fig:6}). The $q$-dependent permittivity has two poles, and exhibits a region of negative values, the "overscreening" effect, in agreement with the results in the literature~\cite{bopp1996}. $\epsilon^{\rm w}$ tends to one, the vacuum response, for large q. The forbidden zone, $\epsilon^{\rm w}\notin [0,1[$, is shaded in red. 
\subsection{Derivation of $\Xi$,  the grand partition function in the grand canonical ensemble}
This subsection presents the main steps for deriving the grand partition function $\Xi$ in the grand canonical ensemble. 
We start from the partition function of the canonical ensemble, $\mathcal{Z}$,
\begin{eqnarray}
\label{Zloc}
\mathcal{Z}=\frac{1}{N_+!}\frac{1}{N_-!}\left[\prod_{i=1}^{N+}\int d\vec{r}_i^+ \right]\left[\prod_{j=1}^{N-}\int d\vec{r}_j^- \right]\int \mathcal{D}[\bm{\mathcal{P}}]e^{-\beta \mathcal{U}_{\rm conf}[\bm{ \mathcal{P}}]}
e^{-\frac{\beta}{2}\int d\vec{r} d\vec{r}'\rho_{\rm tot}(\vec{r})v(\vec{r}-\vec{r}')\rho_{\rm tot}(\vec{r}')}
\end{eqnarray} 
with $v(\vec{r}-\vec{r}')=1/4\pi \epsilon_0|\vec{r}-\vec{r}'|$ the Coulomb potential and $\rho_{\rm tot}(\vec{r})=\rho(\vec{r})-\nabla\cdot\bm{\mathcal{P}}(\vec{r})$.
We introduce an auxiliary field $\Phi$ and perform a Hubbard-Stratonovich transformation using the relation $v(\vec{r}-\vec{r}')^{-1}=-\epsilon_0 \nabla^2 \delta (\vec{r}-\vec{r}')$~\cite{levy2012}. Dropping the prefactor, we obtain
\begin{eqnarray}
\mathcal{Z}=\frac{1}{N_+!}\frac{1}{N_-!}\left[\prod_{i=1}^{N+}\int d\vec{r}_i^+ \right]\left[\prod_{j=1}^{N-}\int d\vec{r}_j^- \right]\int \mathcal{D}[\bm{\mathcal{P}}]e^{-\beta \mathcal{U}_{\rm conf}[\bm{ \mathcal{P}}]}\int \mathcal{D}[\Phi]e^{-\frac{\beta}{2}\int d\vec{r} \epsilon_0 (\nabla \Phi)^2-i\beta \int d\vec{r} \Phi(\vec{r})\left(\rho(\vec{r})-\nabla \cdot\bm{ \mathcal{P}}(\vec{r})\right)}.
\end{eqnarray}
Introducing the expression of the charge density $\rho(\vec{r})=\Sigma_{i=1}^{N+} e \delta(\vec{r}-\vec{r}_i^+)-\Sigma_{j=1}^{N-} e \delta(\vec{r}-\vec{r}_j^-)$, we obtain
\begin{eqnarray}
\mathcal{Z}&=&\int\mathcal{D}[\Phi]\frac{1}{N_+!}\left(\int d\vec{r} e^{-i\beta e \Phi(\vec{r})}\right)^{N_+}\frac{1}{N_-!}\left(\int d\vec{r} e^{i\beta e \Phi(\vec{r})}\right)^{N_-}e^{-\frac{\beta}{2}\int d\vec{r} \epsilon_0(\nabla \Phi(\vec{r}))^2}\nonumber \\
&\times& \int\mathcal{D}[\bm{\mathcal{P}}]e^{-\beta \mathcal{U}_{\rm conf}[\bm{ \mathcal{P}}]} e^{-i \beta\int d\vec{r} \Phi(\vec{r})(-\nabla \cdot \bm{ \mathcal{P}}(\vec{r}))}
\end{eqnarray}
The partition function is brought into a more manageable form by going to the grand canonical ensemble, where we get
\begin{eqnarray}
\Xi&=&\sum_{N+=0}^{\infty}\frac{e^{\beta \mu_+ N_+}}{N_+!}\left(\int d\vec{r} e^{-i \beta e \Phi(\vec{r})}\right)^{N+}
 \times \sum_{N-=0}^{\infty}\frac{e^{\beta \mu_- N_-}}{N_-!}\left(\int d\vec{r} e^{i \beta e \Phi(\vec{r})}\right)^{N-}\nonumber\\
&\times&\int \mathcal{D}[\bm{\mathcal{P}}]e^{-\beta \mathcal{U}_{\rm conf}[\bm{\mathcal{P}}]}
\int \mathcal{D}[\Phi] e^{-\frac{\beta}{2}\int d\vec{r}\epsilon_0\left(\nabla \Phi(\vec{r})\right)^2
	-i \beta\int d\vec{r} \Phi(\vec{r})\left(-\nabla \cdot {\bm{ \mathcal{P}}}(\vec{r})\right)},
\end{eqnarray}
with $\mu_i$, $(i=+,-)$, the chemical potential for cations and anions.  We introduce the field $\Psi=i\Phi$ that can be identified as the electrostatic potential~\cite{levy2012}, we perform the sums over $N_\pm$ and identify the action $F_u[\Psi,\bm{\mathcal{P}}]$, such that
\begin{equation}
\label{GrandPartFunc}
\Xi=\int \mathcal{D}[\bm{\mathcal{P}}]\,\mathcal{D}[\Psi]e^{ - \beta F_u [\bm{{\mathcal{P}}},\Psi]}. 
\end{equation}
Setting $e^{\beta \mu_{\pm}}=n$, the ionic density, we obtainf for the action:
\begin{eqnarray}
\label{action}
F_u[{\bm {\mathcal P}},\Psi]=\mathcal{U}_{\rm conf}[\bm{\mathcal{P}}] -\int d\vec{r}\Big(\frac{\epsilon_0}{2} (\nabla \Psi)^2-\Psi \nabla \cdot \bm{\mathcal{P}}-\frac{2 n}{\beta}\cosh(\beta \Psi e)\Big).
\end{eqnarray}
 
\subsection{Mean field  ($\psi$, P) for the Gaussian limit and the nonlinear model}
In this subsection, we derive  $(\psi, \bm{P})$, the fields minimizing the action $F_u$ given in Eq.~(\ref{action}). $\mathcal{U}_{\rm conf}$ equals to:
\begin{eqnarray}
\label{HnG}
\mathcal{U}_{\rm conf}[\bm{\mathcal{P}}]=\frac{1}{2\epsilon_0}\int d\vec{r}\Big[ \gamma \bm{ \mathcal{P}}(\vec{r})^4+K\bm{\mathcal{P}}(\vec{r})^2
+\kappa_l(\bm{\nabla}\cdot \bm{\mathcal{P}}(\vec{r}))^2+\kappa_t(\bm{\nabla}\times \bm{ \mathcal{P}}(\vec{r}))^2+\alpha (\bm{\nabla}
(\bm{\nabla} \cdot \bm{ \mathcal{P}}(\vec{r})))^2 \Big ].
\end{eqnarray}
The mean fields obey the following equations:
\begin{equation}
\frac{\delta F_u(\vec{P},\psi)}{ \delta \Psi}=0, \quad \frac{\delta F_u(\vec{P},\psi)}{ \delta \mathcal{P}_i}=0, \quad i=x,y,z. 
\end{equation}
The functional derivative with respect to $\Psi$ gives:
\begin{eqnarray}
\frac{\delta F_u}{ \delta \Psi}&=& \epsilon_0 \Delta \Psi -\nabla\cdot \bm{{\mathcal {P}}}-2en  \sinh(\beta e \Psi), 
\end{eqnarray}
The functional derivative with respect to ${\mathcal{P}}_i$ leads to 

\begin{eqnarray}
\frac{\delta F_u}{\delta {\mathcal{P}}_i}=\frac{1}{\epsilon_0}\Big(2\gamma{\mathcal{P}_i}{\bm{\mathcal{P}}}^2+K{\mathcal{P}_i}-\kappa_l\partial_i \nabla \cdot {\bm{\mathcal{P}}} +\kappa_t \left(\partial_i \nabla \cdot {\bm{\mathcal{P}}}-\Delta \mathcal{P}_i\right)+\alpha\partial_i \Delta \nabla \cdot {\bm{\mathcal{P}}} \Big)+\partial_i \Psi.
\end{eqnarray}
The Gaussian limit is obtained by setting $\gamma=0$.

We obtain:
\begin{equation}
	\psi=0, \quad {\bm P}=0.
\end{equation}
The mean fields are vanishing both for Gaussian and nonlinear configuration energy.
 \subsection{Susceptibility in the Gaussian limit ($\gamma$=0) for an electrolyte}
 In this section, we derive the expressions of the longitudinal $\chi_{\parallel}(q)$ and transverse $\chi_\perp(q)$ polarization susceptibility of an electrolyte. \\
 We first define the total susceptibility of the system as the 4x4 matrix
 \begin{eqnarray}
 \label{DefSusc}
\chi_{\rm tot}({\vec{r}_1-\vec{r}_2})= \left(\begin{array}{cc} \epsilon_0\chi & \chi_{P,\psi} \\ \chi_{\psi, P} & \chi_{\psi,\psi}/\epsilon_0
 \end{array}\right)({\vec{r}_1-\vec{r}_2})=\left(\begin{array}{cccc}
 \frac{\delta^2 F_u(\bm{P},\Psi)}{\delta \mathcal{P}_x(\vec{r}_1) \delta \mathcal{P}_x(\vec{r}_2) }  &  \frac{\delta^2 F_u(\bm{P},\Psi)}{\delta \mathcal{P}_x(\vec{r}_1) \delta \mathcal{P}_y(\vec{r}_2) } & \frac{\delta^2 F_u(\bm{P},\Psi)}{\delta \mathcal{P}_x(\vec{r}_1) \delta \mathcal{P}_z(\vec{r}_2) }   & \frac{\delta^2 F_u(\bm{P},\Psi)}{\delta \mathcal{P}_x(\vec{r}_1) \delta \Psi(\vec{r}_1)} \\
 \frac{\delta^2 F_u(\bm{P},\Psi)}{\delta \mathcal{P}_y(\vec{r}_1) \delta \mathcal{P}_x(\vec{r}_2) }  &  \frac{\delta^2 F_u(\bm{P},\Psi)}{\delta \mathcal{P}_y(\vec{r}_1) \delta \mathcal{P}_y(\vec{r}_2) } & \frac{\delta^2 F_u(\bm{P},\Psi)}{\delta \mathcal{P}_y(\vec{r}_1) \delta \mathcal{P}_z(\vec{r}_2) }   & \frac{\delta^2 F_u(\bm{P},\Psi)}{\delta \mathcal{P}_y(\vec{r}_1) \delta \Psi(\vec{r}_1)} \\
 \frac{\delta^2 F_u(\bm{P},\Psi)}{\delta \mathcal{P}_z(\vec{r}_1) \delta \mathcal{P}_x(\vec{r}_2) }  &  \frac{\delta^2 F_u(\bm{P},\Psi)}{\delta \mathcal{P}_z(\vec{r}_1) \delta \mathcal{P}_y(\vec{r}_2) } & \frac{\delta^2 F_u(\bm{P},\Psi)}{\delta \mathcal{P}_z(\vec{r}_1) \delta \mathcal{P}_z(\vec{r}_2) }   & \frac{\delta^2 F_u(\bm{P},\Psi)}{\delta \mathcal{P}_z(\vec{r}_1) \delta \Psi(\vec{r}_1)} \\
 \frac{\delta^2 F_u(\bm{P},\Psi)}{\delta \Psi(\vec{r}_1) \delta \mathcal{P}_x(\vec{r}_2)} &  \frac{\delta^2 F_u(\bm{P},\Psi)}{\delta \Psi(\vec{r}_1) \delta \mathcal{P}_y(\vec{r}_2)} &  \frac{\delta^2 F_u(\bm{P},\Psi)}{\delta \Psi(\vec{r}_1) \delta \mathcal{P}_z(\vec{r}_2)} & \frac{\delta^2 F_u(\bm{P},\Psi)}{\delta \Psi(\vec{r}_1) \delta \Psi(\vec{r}_2)}
 \end{array}\right) ^{-1}
 \end{eqnarray}
 with ($\bm{P},\psi$) denoting the mean fields minimizing the action $F_u$.
We perform the second variational derivatives of $F_u$ as follows:
 \begin{eqnarray}
 \label{F21G}
 \frac{\delta^2 F_u(\bm{P},\psi)}{\delta \Psi(\vec{r}) \delta \Psi(\vec{r}')}&=&\left( \epsilon_0\Delta_\vec{r}-2\Lambda\beta e^2{\rm cosh}(\beta e\psi) \right)\delta(\vec{r}-\vec{r}')\\
 \label{F22G}
 \frac{\delta^2 F_u(\bm{P},\psi)}{\delta \mathcal{P}_i(\vec{r}) \delta \mathcal{P}_j(\vec{r}')}&=&\frac{1}{\epsilon_0}\Big(K\delta_{ij}-\kappa_l \partial_i\partial_j  +\kappa_t\left(\partial_i\partial_j-\Delta \delta_{ij}\right)+\alpha\Delta \partial_i\partial_j \Big)\delta(\vec{r}-\vec{r}'),\quad (i,j)\in(x,y,z)\\
 \label{F23G}
 \frac{\delta^2 F_u(\bm{P},\psi)}{\delta \Psi(\vec{r}) \delta \mathcal{P}_i(\vec{r}')}&=&- \partial_i \delta(\vec{r}-\vec{r}'), \quad
 \frac{\delta^2 F_u(\bm{P},\psi)}{\delta \mathcal{P}_i(\vec{r}) \delta \Psi(\vec{r}')}= \partial_i \delta(\vec{r}-\vec{r}'), \quad i\in(x,y,z).
 \end{eqnarray}

\begin{figure}
	\includegraphics{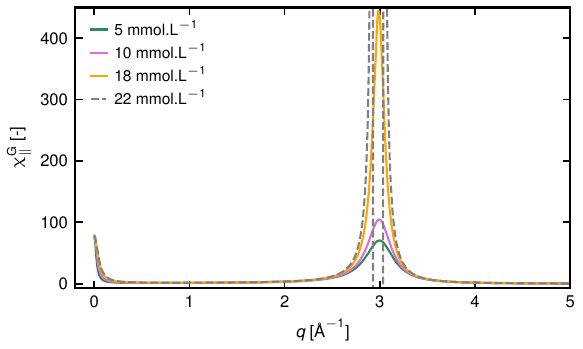}
	\caption{Longitudinal susceptibility of electrolytes. The susceptibility in the Gaussian limit $\chi_\parallel(q)$ given in Eq.~(\ref{chiparaG}) is plotted for increasing salt concentration. Parameters are given in the caption of Fig.~S1.}
	\label{fig:2}
\end{figure}

We thus find for the inverse susceptibility of the system, in Fourier space,
 \begin{eqnarray}
 \label{chiinverse}
 \chi^{-1}_{\rm tot}(\vec{q})=\left(\begin{array}{cccc}
 & & & iq_x \\
 &\frac{\chi^{-1}}{\epsilon_0}(\vec{q})&  & iq_y\\
 & & &iq_z\\
 -iq_x &-iq_y&-iq_z & -(2n e^2\beta+\epsilon_0q^2)
 \end{array}\right),
 \end{eqnarray}
 with the matrix $\chi^{-1}(\vec{q})$ given by
 \begin{eqnarray}
 \label{chim1G}
\chi^{-1}_{ij}(\vec{q})=(K+\kappa_l q^2+\alpha q^4)\frac{q_iq_j}{q^2}+(K+\kappa_t q^2)\left(\delta_{ij}-\frac{q_iq_j}{q^2}\right).   
 \end{eqnarray}
 We invert the matrix $\chi_{\rm tot}^{-1}(q)$ by using a block inversion and obtain
 \begin{eqnarray}
 \chi_{\rm tot}(\vec{q})=\left(\begin{array}{cc} \epsilon_0\chi(\vec{q}) & iQ^{\top} \\
 -i Q & \frac{\chi_{\psi,\psi}(q)}{\epsilon_0}	\end{array}\right)
 \end{eqnarray}
 with $Q=(q_x,q_y,q_z)$ and
 \begin{eqnarray}
 \label{chiG}
 \chi_{ij}(\vec{q})&=&\chi_{\parallel}(q)\frac{q_iq_j}{q^2}+\chi_\perp(q)(\delta_{ij}-\frac{q_iq_j}{q^2})\\
 \label{chiparaG}
 \chi_{\parallel}(q)&=&\frac{\frac{\epsilon_w}{\lambda_D^2}+q^2}{\left(\frac{\epsilon_w}{\lambda_D^2}+q^2\right)(K+\kappa_l q^2+\alpha q^4)+q^2}, \quad
\chi_{\perp}(q)=\frac{1}{K+\kappa_tq^2}\\
 \chi_{\psi,\psi}(q)&=&-\frac{K+\kappa_l q^2+\alpha q^4}{\left(\frac{\epsilon_w}{\lambda_D^2}+q^2\right)(K+\kappa_l q^2+\alpha q^4)+q^2}
 \end{eqnarray}
where we have introduced the Debye length $\lambda_D=\sqrt{\epsilon_0\epsilon_w/2\beta n e^2}$. 
We plot the susceptibility $\chi_{\parallel}(q)$ for increasing salt concentrations ($c=n/\mathcal{N}_a$, $\mathcal{N}_a$ the Avogadro number) in Fig.~\ref{fig:2}. As we have ($\kappa_l<$0, $\alpha>$0), the denominator of $\chi_{\parallel}(q)$ will vanish for a small enough Debye length inducing a divergence of the susceptibility. For the chosen set of parameters, the divergence occurs for $c$=22 mmol.l$^{-1}$. At this concentration, the longitudinal correlations are purely oscillating, illustrating an unphysical crystalization of the medium.
  \subsection{Susceptibility for Gaussian model in real space}
\begin{figure}
 	\includegraphics[scale=0.27]{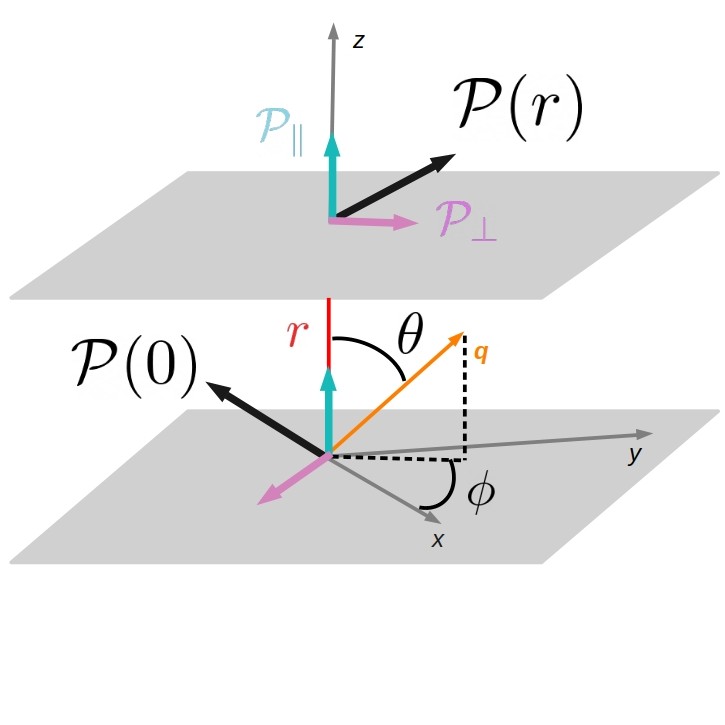} 
 	\caption{Illustration of the basis considered to calculate the susceptibility tensor $\chi$.}
 	\label{fig:3}
 \end{figure}

We express the Gaussian susceptibility $\chi$ given in Eq.~(\ref{chiparaG}) in real space.  We perform the back Fourier transform   $\hat{\chi}_{ij}(\vec{r})$=$\mathcal{F}_T(\chi_{ij}(\vec{q}))$ defined as
 \begin{equation}
 \label{chifourier}
 \hat{\chi}_{ij}(\vec{r})=\frac{1}{2\pi^3}\int_0^{2\pi} d\phi \int_0^{\pi} d\theta \sin(\theta) \int_0^\infty dq q^2 e^{iqr\cos(\theta)}\chi_{ij}(\vec{q}).
 \end{equation}
$\hat{\chi}$ is here expressed in the intrinsic basis in which the vector $\vec{r}$, joining the two correlated points, is aligned with the $\vec{e}_z$ direction of wavenumber $\vec{q}$ basis, as illustrated in Fig.~\ref{fig:3}.
 We express the longitudinal projector $q_iq_j/q^2$ $(i,j=x,y,z)$ in the  spherical basis, 
 \begin{equation}
 \label{longq}
 \frac{q_iq_j}{q^2}=\left(\begin{array}{ccc}
 \sin(\theta)^2\cos(\phi)^2 & 	\sin(\theta)^2\cos(\phi)\sin(\phi) & 	\sin(\theta)\cos(\theta)\cos(\phi) \\
 \sin(\theta)^2\cos(\phi)\sin(\phi) & \sin(\theta)^2\sin(\phi)^2 & 	\sin(\theta)\cos(\theta)\sin(\phi) \\
 \sin(\theta)\cos(\theta)\cos(\phi) & 	\sin(\theta)\cos(\theta)\sin(\phi) & \cos(\theta)^2
 \end{array}\right),
 \end{equation}
Using Eqs.~(\ref{chiG},\ref{chiparaG}), $\hat{\chi}_{ij}(r)$ is split into two contributions: $\hat{\chi}_{ij}(r)=\mathcal{F}_T(\chi_{\parallel}(q)q_iq_j/q^2)+\mathcal{F}_T(\chi_{\perp}(q)(\delta_{ij}-q_iq_j/q^2))$. We perform the integrals in Eq.~(\ref{chifourier}) using (\ref{longq}). We obtain:
 \begin{equation}
 \label{spericcorrelation}
 \mathcal{F}_T(\chi_{\parallel}(q)q_iq_j/q^2)= \left(\begin{array}{ccc}\frac{1}{2}(I_{1,\parallel}(r)-I_{2,\parallel}(r)) &0& 0\\
 0 & \frac{1}{2}(I_{1,\parallel}(r)-I_{2,\parallel}(r))  & 0 \\
 0 & 0 & I_{2,\parallel}(r)\end{array}\right),
 \end{equation}
and

 \begin{eqnarray}
\mathcal{F}_T\left(\chi_{\perp}(q)(\delta_{ij}-q_iq_j/q^2)\right)=\left(\begin{array}{ccc}
 \frac{1}{2}\left(I_{1,\perp}(r) + I_{2,\perp}(r) \right) &0 &\\0&	\frac{1}{2}\left(I_{1,\perp}(r)+ I_{2,\perp}(r)\right) &0 \\
 0 & 0 & 	I_{1,\perp}(r) -I_{2,\perp}(r) 
 \end{array}\right).
 \end{eqnarray}
We have introduced four elementary functions ($I_{\parallel,1}$, $I_{\perp,1}$ $I_{\parallel,2}$, $I_{\perp,2}$), defined as follow:
 \begin{eqnarray}
 \label{Iexp}
 	I_{i,1}(r)&=&\frac{1}{(2 \pi)^3}\int_0^\infty dq q^2 \int_0^\pi d\theta \sin(\theta) \int_0^{2\pi} d\phi\, \chi_i(q)e^{iqr\cos(\theta)},\nonumber\\ \quad I_{i,2}(r)&=&\frac{1}{(2 \pi)^3}\int_0^\infty dq q^2\int_0^\pi d\theta\sin(\theta)\cos(\theta)^2 \int_0^{2\pi}  d\phi\, \chi_i(q)e^{iqr\cos(\theta)}, 
 \end{eqnarray}
with $\quad i=\parallel,\perp$.
  The susceptibility finally reads:
 \begin{eqnarray}
 \hat{\chi}(r)=\left(\begin{array}{ccc}
\hat{\chi}_{\perp}(r) & 0 & 0\\ 
 0 &\hat{\chi}_{\perp}(r) & 0\\
 0 &0 & \hat{\chi}_{\parallel}(r) 
 \end{array}\right)
 \end{eqnarray}
with 
 \begin{eqnarray}
 \label{chidiagr}
 \hat{\chi}_{\parallel}(r)&=&I_{2,\parallel}(r)+(I_{1,\perp}(r)-I_{2,\perp}(r)) \\
 \label{chidiagangl}
 \hat{\chi}_{\perp}(r)&=&\frac{I_{1,\parallel}(r)-I_{2,\parallel}(r)}{2}+\frac{I_{1,\perp}(r)-I_{2,\perp}(r)}{2},
 \end{eqnarray}
 in the basis $\{\vec{e}_i\}$, $(i=x,y,z)$, defined such that $\vec{r}$ is aligned with  $\vec{e}_z$, see sketch in Fig.~\ref{fig:3}. \par
The susceptibility associated with a distance vector $\vec{r}=(x,y,z))$ is  obtained by performing the following change of basis:
\begin{equation}
\label{chicart}
	\hat{\chi}_{\rm cart}=R^{-1}\cdot \hat{\chi}(r)\cdot R,
\end{equation}
$R$ is the change-of-basis matrix from Cartesian to spherical coordinates.
\subsection{One-loop free energy expansion}
In the case of nonvanishing $\gamma$, the configurational integrals cannot be performed exactly. The partition function can be expanded to the second order around the mean field point $(\bm{P},\psi)$, as
\begin{eqnarray}
\Xi&\approx&e^{-\beta F_u[\bm{P},\psi]}\int \mathcal{D}[\delta\bm{\mathcal{P}}]\, \mathcal{D}[\delta\Psi]e^{-\frac{\beta}{2}\int d\vec{r}d\vec{r}'\left(\delta \bm{\mathcal{P}}(\vec{r}), \delta\Psi(\vec{r}) \right)\cdot F_u^{(2)}({\bm P},\psi) \cdot \left(\delta \bm{\mathcal{P}}(\vec{r}), \delta\Psi(\vec{r}) \right) }\nonumber\\&=&e^{-\beta F_u[\psi,P]}\left(\beta |F_u^{(2)}|\right)^{-1/2}
\end{eqnarray}
where we have dropped the prefactor.
$F_u^{(2)}({\bm r}-{\bm r}')$ is the second functional derivative of the action and is defined as follows:
\begin{eqnarray}
F_u^{(2)}(\vec{r}-\vec{r}')=\left(\begin{array}{cc}
\frac{\delta^2 F_u(\bm{P},\psi)}{\delta \mathcal{P}_i(\vec{r}) \delta \mathcal{P}_j(\vec{r}')}    & \frac{\delta^2 F_u(\bm{P},\psi)}{\delta \mathcal{P}_i(\vec{r}) \delta \Psi(\vec{r}')} \\
\frac{\delta^2 F_u(\bm{P},\psi)}{\delta \Psi(\vec{r}) \delta \mathcal{P}_i(\vec{r}')}  & \frac{\delta^2 F_u(\bm{P},\psi)}{\delta \Psi(\vec{r}) \delta \Psi(\vec{r}')}
\end{array}\right)
\end{eqnarray}
with
\begin{eqnarray}
\label{F21}
\frac{\delta^2 F_u(\bm{P},\psi)}{\delta \Psi(\vec{r}) \delta \Psi(\vec{r}')}&=&\left( \epsilon_0\Delta_\vec{r}-2\Lambda\beta e^2{\rm cosh}(\beta e\psi) \right)\delta(\vec{r}-\vec{r}')\\
\label{F22}
\frac{\delta^2 F_u(\bm{P},\psi)}{\delta \mathcal{P}_i(\vec{r}) \delta \mathcal{P}_j(\vec{r}')}&=&\frac{1}{\epsilon_0}\Big(\left(2\gamma P_{i}^2+K\right)\delta_{ij}+4\gamma P_i P_j-\kappa_l \partial_i\partial_j  +\kappa_t\left(\partial_i\partial_j-\Delta \delta_{ij}\right)+\alpha\Delta \partial_i\partial_j \Big)\delta(\vec{r}-\vec{r}')\\
\label{F23}
\frac{\delta^2 F_u(\bm{P},\psi)}{\delta \Psi(\vec{r}) \delta \mathcal{P}_i(\vec{r}')}&=&- \partial_i \delta(\vec{r}-\vec{r}'), \quad
\frac{\delta^2 F_u(\bm{P},\psi)}{\delta \mathcal{P}_i(\vec{r}) \delta \Psi(\vec{r}')}= \partial_i \delta(\vec{r}-\vec{r}')
\end{eqnarray}
 The free energy at first in the loop expansion follows: 
\begin{equation}
\label{FreeEnergyOneLoop}
\mathcal{F}\approx F_u[{\bm P},\psi]+\frac{1}{2\beta}{\rm Tr}{\rm ln}{\beta F_u^{(2)}}[{\bm P},\psi].
\end{equation}

 \subsection{One-loop correction for the inverse susceptibility}
 In this subsection, we expand the inverse polarization susceptibility to the first loop order, $\chi^{-1}\approx \chi_{(0)}^{-1}+\chi_{(1)}^{-1}$. 
Using its expression as a function of the free energy of the system,
 \begin{eqnarray}
 \label{DefSusc}
 \frac{1}{\epsilon_0}\chi^{-1}_{ij}(\vec{r}_1-\vec{r}_2)=
 \frac{\delta^2 \mathcal{F}(\bm{P},\psi)}{\delta \mathcal{P}_i(\vec{r}_1) \delta \mathcal{P}_j(\vec{r}_2) }
 \end{eqnarray}
 and the first order expansion of $\mathcal{F}$, (Eq.  \ref{FreeEnergyOneLoop}), we obtain:
 \begin{eqnarray}
 \label{MatrDeriv}
\frac{1}{\epsilon_0}(\chi_{(1)}^{-1})_{xx}(\vec{r}_1-\vec{r}_2) & =&\frac{\delta {\rm Tr} {\rm ln} (\beta F_u^{(2)})}{\delta \mathcal{P}_x(\vec{r}_1)\delta \mathcal{P}_x(\vec{r}_2)}(\bm{P},\psi)\nonumber \\ &=&{\rm Tr}\left( \left(F_u^{(2)}\right)^{-1} \cdot \frac{\partial^2F_u^{(2)} }{\partial \mathcal{P}_x(\vec{r}_1)\partial \mathcal{P}_{x}(\vec{r}_2)}-\frac{\delta F_u^{(2)} }{\delta \mathcal{P}_x(\vec{r}_1)}\frac{\delta F_u^{(2)} }{\delta \mathcal{P}_x(\vec{r}_2)}\cdot \left(F_u^{(2)}\right)^{-2}\right)\Big(\bm{P},\psi\Big) .
 \end{eqnarray}
 The second matrix product in the right-hand term is vanishing as $\delta F_u^{(2)} / \delta \mathcal{P}_x(\vec{r}_2)(\bm{P},\psi)=0$. We thus obtain:
 \begin{equation}
 \label{MatrDerivxx0}
 \frac{\delta {\rm Tr} {\rm ln} (\beta F_u^{(2)})}{\delta \mathcal{P}_x(\vec{r}_1)\delta \mathcal{P}_{x}(\vec{r}_2)}(\bm{P},\psi) ={\rm Tr}\left( \left(F_u^{(2)}\right)^{-1} \cdot \frac{\partial^2F_u^{(2)} }{\partial \mathcal{P}_x(r_1)\partial \mathcal{P}_{x}(r_2)}\right)(\bm{P},\psi) .
 \end{equation}
We use the expression of $F_u^{(2)}$ given in Eq. (\ref{F22}) and obtain
 \begin{equation}
 \label{MatrDerivxx}
 \frac{\partial^2F_u^{(2)}(\bm{P},\psi) }{\partial \mathcal{P}_x(r_1)\partial \mathcal{P}_x(r_2)}=	\left(\begin{array}{cccc} 12 \gamma /\epsilon_0 & 0 &0 & 0 \\
 0 & 4  \gamma /\epsilon_0& 0 & 0 \\
 0 & 0 & 4  \gamma /\epsilon_0 & 0 \\
 0 & 0 & 0 & 0
 \end{array}\right)\delta(\vec{r}-\vec{r}_1)\delta(\vec{r}-\vec{r}_2)\delta(\vec{r}'-\vec{r})
 \end{equation}
 and 
 \begin{equation}
 \label{d2F2}
 \frac{\partial^2F_u^{(2)}(\bm{P},\psi) }{\partial P_x(\vec{r}_1)\partial P_y(\vec{r}_2)}=\left(\begin{array}{cccc} 0 & 4 \gamma/\epsilon_0 &0 & 0 \\
 4 \gamma/\epsilon_0 & 0 & 0 & 0 \\
 0 & 0 & 0 & 0 \\
 0 & 0 & 0 & 0
 \end{array}\right)\delta(\vec{r}-\vec{r}_1)\delta(\vec{r}-\vec{r}_2)\delta(\vec{r}'-\vec{r})
 \end{equation}
 The other matrices are easily deduced by symmetry.

 We now have to calculate the trace of the matrices in (\ref{MatrDerivxx}) by integrating over the continuous indices ${\int d\vec{r}\,d\vec{r}'}$ and summing over the discrete indices. 
 The polarization correlations depend only on the distance $u=|\vec{r}-\vec{r}'|$.
 We thus write:
 \begin{eqnarray}
\int d\vec{r} d\vec{r}' \left(F_u^{(2)}\right)^{-1} \cdot \frac{\partial^2F_u^{(2)} }{\partial \mathcal{P}_i(\vec{r}_1)\partial \mathcal{P}_j(\vec{r}_2)}= \int d\vec{r}\int du u^2\left( \int d\phi\, d\theta\sin(\theta)  \left(F_u^{(2)}\right)^{-1} \right) \frac{\partial^2F_u^{(2)} }{\partial \mathcal{P}_i(\vec{r}_1)\partial \mathcal{P}_j(\vec{r}_2)}. 
 \end{eqnarray}  
Using $\delta (\vec{r}-\vec{r}')=\delta(u)/2 \pi u^2$ and $\Big(F_u^{(2)}\Big)^{-1}(\bm{P},\psi)=\chi_{\bm{P},\psi}(\vec{r}-\vec{r}')$ given in Eq.~(\ref{DefSusc}),  
 we perform the integral over $\theta$ and $\phi$ and find for the polarization correlation:
 \begin{eqnarray}
 \label{chirbasis}
 \int_0^{2\pi}d\phi\int_0^{\pi}d\theta \sin(\theta) R^{-1}\cdot\chi_{ij}(u)\cdot R=\frac{4\pi}{3}\left(\chi_\parallel(u)+2\chi_{\perp}(u)\right)\delta_{ij} .
\end{eqnarray}
 We now evaluate Eq.(\ref{MatrDerivxx0}) and find: 
 \begin{eqnarray}
 \label{chim11}
 (\chi_{(1)}^{-1})_{xx}(\vec{r}_1-\vec{r}_2)&=&	\frac{\pi}{\beta}\int d\vec{r} \int_0^{\infty} du 4 \gamma\epsilon_0\left(\frac{10}{3} \hat{\chi}_\parallel(u)+\frac{20}{3}\hat{\chi}_\perp(u)\right)\frac{\delta(u)}{2\pi}\delta(\vec{r}-\vec{r}_1)\delta(\vec{r}-\vec{r}_2)\nonumber\\
 \label{finalres}
 &=&\frac{20 \gamma\epsilon_0}{ 3 \beta }\left(\hat{\chi}_\parallel(0)+2\hat{\chi}_\perp(0)\right)\delta(\vec{r}_1-\vec{r}_2).
 \end{eqnarray}
 Note that we find the same value for $(\chi_{(1)}^{-1})_{yy}$ and  $(\chi_{(1)}^{-1})_{zz}$  and that the cross terms are vanishing. We calculate the susceptibility at $r=0$, $\chi(0)$, using the  elementary functions defined in Eq.~(\ref{Iexp}) as follows,
 \begin{eqnarray}
 \label{i1para}
 	I_{1,\parallel}(0)&=&\frac{1}{2\pi^2}\int_0^\infty dq q^2\frac{\epsilon_w/\lambda_D^2+q^2}{(\epsilon_w/\lambda_D^2+q^2)(K+\kappa_lq^2+\alpha q^4)+q^2},\\
 	\label{i1perp}
 	I_{1,\perp}(r_c)&=&\frac{1}{2\pi^2}\int_0^{2\pi/r_c}dq\frac{q^2}{K+\kappa_t q^2}=\frac{1}{\pi K \lambda_t^2r_c}\\
 	\label{i2}
 	I_{2,\parallel}(0)&=&\frac{I_{1,\parallel}(0)}{3}, \quad I_{2,\perp}(r_c)=\frac{I_{1,\perp}(r_c)}{3}
\end{eqnarray}
  where we have introduced a cutoff length $r_c$ to remove the divergence for $I_{1/2,\perp}$ in $r=0$. \par  The Gaussian susceptibility tensor reads,
  \begin{equation}
  \label{chirm}
  	\hat{\chi}_{\parallel}(r_c)=I_{2,\parallel}(0)+I_{1,\perp}(r_c)-I_{2,\perp}(r_c),\quad \hat{\chi}_{\perp}(r_c)=\frac{1}{2}\left(I_{1,\parallel}(0)-I_{2,\parallel}(0)+I_{1,\perp}(r_c)-I_{2,\perp}(r_c)\right).
  \end{equation} 
 Using Eq. (\ref{finalres}), we obtain in Fourier space,
 \begin{equation}
 \label{deltaK}
 	\chi^{-1}_{(1)}(\vec{q})=\delta K\frac{q_iq_j}{q^2}+\delta K\left(\delta_{ij}-\frac{q_iq_j}{q^2}\right),\quad {\rm with}\quad \delta K=\frac{20\gamma\epsilon_0}{3\beta}\left(\hat{\chi}_{\parallel}(r_c)+2\hat{\chi}_{\perp}(r_c)\right).
 \end{equation}
 Finally, we obtain the expression for the inverse polarization susceptibility to first order,
 \begin{eqnarray}
 	\chi_{ij}^{-1}(\vec{q})=(K+\delta K+\kappa_l q^2+\alpha q^4)\frac{q_iq_j}{q^2}+(K+\delta K+\kappa_t q^2)\left(\delta_{ij}-\frac{q_iq_j}{q^2}\right).	
 \end{eqnarray}
 where we have used the expression of $\chi(\vec{q})$ given in Eq. (\ref{chim1G}).
  \subsection{Linear dependence of $\chi^{-1}_{(1)}(q)$ in salt concentration $c$}
The correction $\delta K$ is split into two contributions, 
\begin{equation}
	\delta K=\delta K_w +\delta K_c c+\tau(c^2)
\end{equation}  
a pure water one $\delta K_w$, and a second one, $\delta K_c c$, depending on the salt concentration and expanded linearly in $c$.

To obtain explicit expressions of $\delta K_w$ and $\delta K_c$, we expand the functions $I_{i,x}(0)$, $i=1,2$, $x=\parallel, \perp$ given in Eq.~(\ref{i1para}-\ref{i2}) linearly in $c$, using $\epsilon_w/\lambda_D^2$=$c\times2\mathcal{N}_ae^2\beta/\epsilon_0$. We obtain
\begin{eqnarray}
	I_{i,\parallel}(0)&=&	I^w_{i,\parallel}(0)+c\times I^1_{i,\parallel}(0)+\tau(c^2)\\
		I_{i,\perp}(r_c)&=&	I^w_{i,\perp}(r_c), \quad i=1,2
\end{eqnarray}
where the functions are split into a pure water contribution and a linear correction in $c$. Note that the $I_{i,\perp}$ functions do not depend on the salt concentration, the associated salt correction is thus vanishing. \\ 
The functions for pure water obey
\begin{eqnarray}
I^w_{1\parallel}(0)&=&\frac{1}{2\pi^2}\int_0^\infty dq q^2\frac{1}{1+K+\kappa_l q^2 +\alpha q^4},\quad 	I^w_{2\parallel}(0)=\frac{I_{1,\parallel(0)}^w}{3},\\
I^w_{1\perp}(r_c)&=&\frac{1}{2\pi^2}\int_0^{\frac{2\pi}{r_c}} dq q^2\frac{1}{K+\kappa_t q^2 },\quad 	I^w_{2\perp}(r_c)=\frac{I_{1,\perp}^w(r_c)}{3},
\end{eqnarray}
and the linear correction for the parallel function reads
\begin{equation}
\label{i1salt}
	I^1_{1,\parallel}(0)=\frac{1}{2\pi^2\epsilon_0}\int_0^\infty dq \frac{2\mathcal{N}_ae^2\beta}{(1+K+\kappa_lq^2+\alpha q^4)^2}, \quad I^1_{2,\parallel}(0)=\frac{I^1_{1,\parallel}(0)}{3}. 
\end{equation}
The expressions of $\hat{\chi}_{\parallel}^{\rm w}(r_c)$ and $\hat{\chi}_{\perp}^{\rm w}(r_c)$ are obtained by replacing $I_{i, \parallel}(0)$ by $I^w_{i, \parallel}(0)$ and $I_{i, \perp}(0)$ by $I^{\rm w}_{i, \perp}(0)$ in Eq.~(\ref{finalres}). Finally, we obtain:
\begin{equation}
	\delta K_w=\frac{20\gamma\epsilon_0}{3\beta}\left(\hat{\chi}^{\rm w}_{\parallel}(0)+\hat{\chi^{\rm w}}_{\perp}(r_c)\right).
\end{equation}
We set $\delta K_w$ to zero as it is included in the fitted parameter $K$. Using the expressions for $\delta K$ in Eq.~(\ref{deltaK}) and ($\hat{\chi}_{\perp}$, $ \hat{\chi}_{\parallel}$) in Eq.~(\ref{chirm}), we obtain
\begin{equation}
	\delta K_c=\frac{20\gamma\epsilon_0}{3\beta}I^1_{1,\parallel}(0).
\end{equation}
To obtain more physical insight, we express the function $I^1_{1,\parallel}(0)$ as a function of the characteristic lengths of the problem.
The longitudinal susceptibility $\chi^{\rm w}_{\parallel}(q)$  is associated with two correlation lengths: a longitudinal decay, $\lambda_d$ and an oscillation length $\lambda_o$, defined as the imaginary and real part of the inverse of the poles of the function. The transverse susceptibility $\chi^{\rm w}_{\perp}(q)$  is associated with a decay length $\lambda_t$
which is the inverse of its pole. We can express these lengths as functions of the parameters of the model. Their expressions obey:
\begin{equation}
\label{longueur}
\lambda_d=\frac{2\sqrt{\alpha}}{\sqrt{2\sqrt{\alpha(1+K)}+\kappa_l}}, \quad \lambda_t=\sqrt{\frac{\kappa_t}{K}}, \quad \lambda_o=\frac{4\pi\sqrt{\alpha}}{\sqrt{2\sqrt{\alpha(1+K)}-\kappa_l}}.
\end{equation}
Using the estimated values of the parameters, we get a longitudinal decay length $\lambda_d$=4.7~\AA, an oscillating length $\lambda_o$=2.1~\AA, and a transverse decay length, $\lambda_t$=1.05\AA.\par
\begin{figure}
	\includegraphics{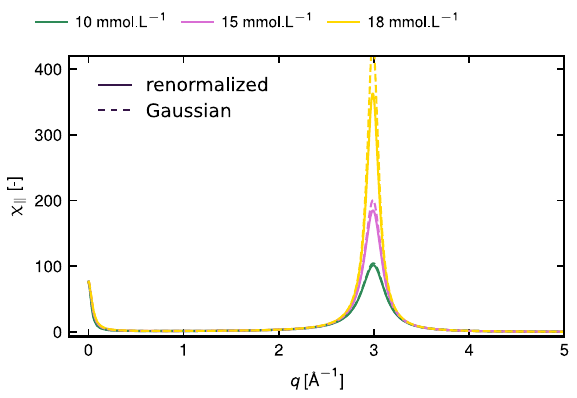}
	\caption{Renormalized longitudinal susceptibility compared with the Gaussian predictions. The susceptibility $\chi_\parallel(q)$ given in Eq. (\ref{oneloopchi}) is plotted for increasing concentration with $\delta K_c$=0.028 mol$^{-1}$.L (solid lines) and compared to the Gaussian susceptibility given in Eq.~(\ref{chiparaG}) (dashed lines). Other parameters are given in the caption of Fig.~S1.}
	\label{fig:4}
\end{figure}

We perform the integral in Eq.~(\ref{i1salt}) and obtain
\begin{equation}
	I_{1,\parallel}^1(0)=\frac{\beta \mathcal{N}_a e^2(\epsilon_w-1)^2}{64\pi \epsilon_0\epsilon_w}\frac{(4\pi^2\lambda_d^2+\lambda_o^2)(4\pi^2\lambda_d^2+5\lambda_o^2)}{\lambda_d\lambda_o^4}
\end{equation}
after having expressed $K$, $\kappa_l$, $\alpha$ as functions of $\epsilon_w$, $\lambda_o$, $\lambda_d$ by inverting Eq.~(\ref{longueur}).
We deduce that,
\begin{equation}
	\delta K_c=\gamma\frac{5 \mathcal{N}_a e^2(\epsilon_w-1)^2}{24\pi \epsilon_w}\frac{(4\pi^2\lambda_d^2+\lambda_o^2)(4\pi^2\lambda_d^2+5\lambda_o^2)}{\lambda_d\lambda_o^4}
\end{equation}
Expanding $\epsilon(c)=1+1/(K+\delta K_c c)$ to linear order in $c$, we obtain 
for the permittivity of the electrolytes
\begin{equation}
	\epsilon(c)=\epsilon_w-\frac{\delta K_c}{K^2}c.
\end{equation}
The inverse susceptibility can thus be written as:
 \begin{eqnarray}
\chi^{-1}(q)&=&\chi^{-1}_{(0)}(q)+\chi^{-1}_{(1)}(q)\nonumber\\
&=&(K+\delta K_c c+\kappa_l q^2+\alpha q^4)\frac{q_iq_j}{q^2}+(K+\delta K_cc+\kappa_t q^2)\left(\delta_{ij}-\frac{q_iq_j}{q^2}\right).	
\end{eqnarray}
Using Eq.~(\ref{chiinverse}) and following the derivation detailed in S2.3, we find:
\begin{equation}
\label{oneloopchi}
	\chi_\parallel(q)=\frac{\frac{\epsilon_w}{\lambda_D^2}+q^2}{\left(\frac{\epsilon_w}{\lambda_D^2}+q^2\right)(K+\delta K_c c+\kappa_l q^2+\alpha q^4)+q^2}.
\end{equation}
We plot the renormalized susceptibility $\chi_\parallel(q)$ for increasing salt concentration in Fig.~\ref{fig:4} (solid lines). By comparing the susceptibilities obtained for the
Gaussian model (dashed lines), we see that the enhancement of the
pseudo-resonant peak at $q$=3~\AA$^{-1}$ is attenuated but not canceled by the one-loop correction.

\medskip
\bibliographystyle{unsrt}
\bibliography{RPfinal}